\documentclass[aps,twocolumn,prb]{revtex4}
\makeatletter
\def\@dotsep{4.5}
\makeatother

\usepackage{graphicx}
\usepackage{amssymb}
\usepackage{natbib}
\usepackage{amsmath}

\newcommand{\comment}[1]{}

\begin{document}

\title{Boron Nitride Nanotubes as Templates for Half-Metal Nanowires}

\author{ Ronaldo J. C. Batista,$^1$}  
\author{Alan B. de Oliveira,$^1$}  
\author{Nat\'alia R. Pereira,$^2$} 
\author{Rafael S. Paolini,$^2$} 
\author {Ta\'{i}se M. Manhabosco$^1$}
\affiliation{$^1$ Departamento de F\'{\i}sica, 
Universidade Federal de Ouro Preto,  Ouro Preto, MG, 35400-000, Brazil.\\ 
$^2$ Escola de Minas, Universidade Federal de Ouro Preto,  Ouro Preto, MG, 35400-000, Brazil.}

\begin{abstract}
We investigate by means of DFT/GGA+U calculations the electronic and structural properties of magnetic nanotubes composed of an iron oxide monolayer
and (n,0) Boron Nitride (BN) nanotubes, with n ranging from 6 up to 14. 
The formation energy per FeO molecule of FeO covered tubes is smaller than the formation energy of small FeO nanoparticles 
which suggest that the FeO molecules may cover the BN nanotubes rather than to aggregate to form
the FeO bulk. We propose a continuous model for the FeO covered BN nanotubes formation energy which
predicts that BN tubes with diameter of roughly 13 \AA~ are the most stable.   
Unlike carbon nanotubes, the band structure of FeO covered BN nanotubes can not be obtained by slicing the
band structure of a FeO layer, the curvature and the interaction with the BN tube is determinant
for the electronic behavior of FeO covered tubes. As a result the tubes are semiconductors, intrinsic half-metals or semi-half-metals 
that can become half-metals charged with either electrons and holes. Such a result may be important in the spintronics context. 

\end{abstract}

\maketitle

\section{Introduction}

Metal oxides are ubiquitous in nature.
Their combination with nanotubes may lead to materials which present several novel properties. Besides the pure scientific interest that lie on those systems, 
there is as well a huge potential for technological applications based on them. Since 1995 it is possible to use nanotubes as templates for metal oxides nanocomposites and nanostructures, as it was shown by Ajayan and collaborators.\cite{ASRC1995} The combination of nanotubes with magnetic oxides may lead to magnetic nanowires which could be used as tips for magnetic scanning microscopes or to control  the spin polarization of electrons, which is very important in the spintronic context. Indeed, it has been shown that iron oxides nanotubes are half-metals\cite{LZHLLLFZ2004} that can be used as core material for spin-filter devices.\cite{LLXZXY2005}

\begin{figure}
 \centering
\includegraphics[scale=0.35,clip=true]{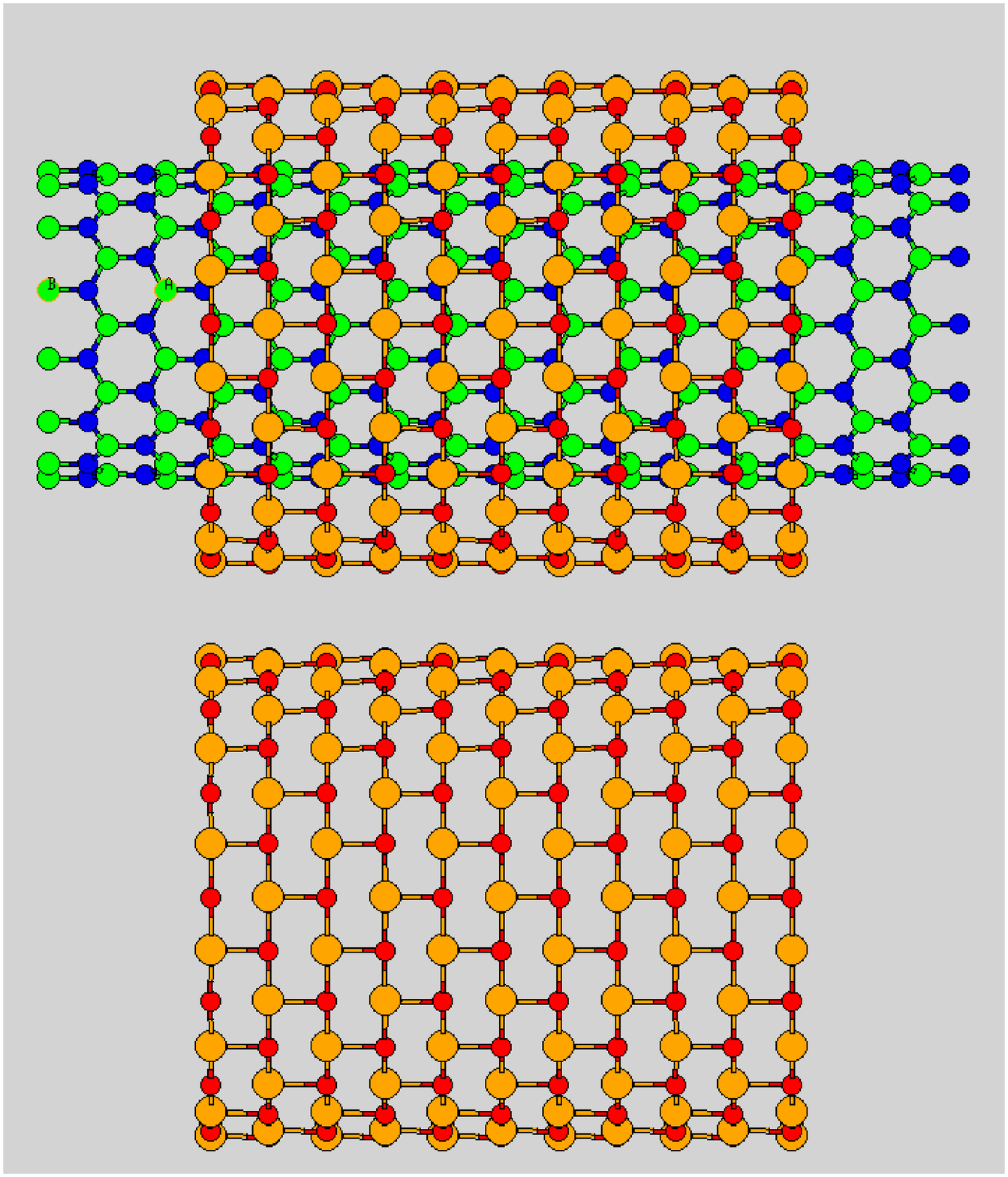}
 \caption{Structure of FeO covered boron nitride (BN) nanotube. The FeO molecules form a square lattice (bottom panel) 
which commensurates with the BN hexagonal lattice along the tubes axis. The Fe-O and B-N bond lengths in the optimized structures (approximately 2.13 and 1.42 \AA, respectively) are very close to the values of bond lengths of FeO bulk (2.15 \AA) 
and hexagonal boron nitride (1.45 \AA),\cite{BOR2010} respectively.}      
 \label{Lateral}
\end{figure}

Boron nitride (BN) structures are insulator and non-magnetic. In this sense it is reasonable to consider BN nanotubes (instead of carbon nanotubes) as the ideal template for half-metal nanowires.
In previous works\cite{BMC2007,BOR2010} it has been shown that FeO molecules strongly interacts with BN structures, where the Fe and O atoms bind to the N and B  atoms through covalent
bonds, respectively. 
In particular, such an interaction is stronger in sites where the local curvature of BN structures is larger. In fact, the tube curvature is essential to  match the hexagonal lattice of 
the boron nitride nanotube with the square lattice of an FeO layer. Besides the curvature,  it is also necessary that the unit cell of the two lattices commensurate along the tube axes in order
to have the two lattices matched. It is important to stress that due to the values of the Fe-O and B-N bond lengths (2.15 and 1.45 \AA~ for the FeO bulk and the hexagonal BN, respectively\cite{BOR2010}) the unit cell of $\emph{zigzag}$ BN nanotubes commensurates, within an error of 1$\%$, with
the unit cell of a square FeO layer, Fig. \ref{Lateral}. This allows one to built up FeO covered BN tubes with the same number of Fe, O, N and B atoms. 

In this work we apply fist-principles formalism to investigate the effect of the tube diameter on the stability and on the electron transport properties
of FeO covered BN nanotubes. We have observed that due the effect of the curvature, the interaction between the FeO layer and BN tubes and the quantization of wave vectors,
the electronic structure is diameter's dependent. 
As a result, a rich variety of electron transport properties were found: (i) the (6,0) tube is semiconductor; (ii) the (10,0) is a half-metal;
(iii) the (11,0) is a semi-half-metal which could become a half-metal if doped with holes; (iv) the  (14,0) is a semi-half-metal which
could become a half-metal if doped with electrons. Regarding the structural properties, we have found that the tube diameter determines  the distance between the FeO layer and the BN tube. This happens because the interaction between 
the BN nanotube and the FeO layer is weaker than the Fe-O and B-N bonds. The formation energy per FeO molecule of FeO covered tubes is smaller than the formation energy of small FeO nanoparticles  which suggests that the FeO molecules may cover the BN nanotubes rather than to aggregate to form
the FeO bulk.  We also propose a continuous model for the FeO covered BN nanotubes formation energy which
predicts that BN tubes with diameter of roughly 13 \AA~ are the most stable.

\section{Methodology}

Our  first-principles methodology is based
on the Density Functional Theory (DFT) as implemented
in the SIESTA program.\cite{siesta} We used the Generalized Gradient
Approximation (GGA) as parametrized in the Perdew-Burke-Ernzerhof scheme (PBE)\cite{pbe96} for 
the exchange-correlation functional. The ionic core potentials were represented
by norm-conserving scalar relativistic Troullier-Martins\cite{martins} pseudopotentials 
in Kleinman-Bylander nonlocal form.\cite{kleinman}
The fineness of the real-space grid integration was defined by a minimal energy
cutoff of 150~Ry.\cite{josemesh} A $8\times1\times1$ Monkhorst-Pack grid was
used to sample the Brillouin zone. 
 Geometries were fully optimized using the conjugate gradient
algorithm\cite{payne} until all 
the force components were smaller than 0.05 eV/\AA.  The Khon-Sham (KS) eigenfunctions were
expanded as linear 
combination of pseudo atomic orbitals of finite range consisting of double-zeta
radial 
functions per angular momentum plus polarization orbitals (DZP). The range of each
atomic orbital 
was determined by a common confinement energy-shift of
$\delta E=0.01$~Ry. \cite{emiliobase}

We have also applied a GGA+U methodology to investigate the electronic properties of FeO covered nanotubes.  The use such a  methodology 
was motivated by its ability to reproduce the observed insulating behavior
for FeO bulk. The GGA approach describes accurately the structural properties of FeO bulk but it tends to over-delocalize $3d$ electrons, which leads to unphysical metallic behavior.

The current GGA+U implementation in the Siesta code is based on the formulation of Dudarev and collaborators.\cite{Dudarev98}
Slightly-excited numerical atomic orbitals of finite range  were used to calculate the local populations used in a Hubbard-like term that modifies the Hamiltonian and energy. The radii of Fe $3d$ and $4s$ orbitals are respectively 2.90 and 1.22~\AA. The effective Hubbard-like terms used in the calculation are 4.65 and 0.95, for the $3d$ and $4s$ shells respectively. 

\section{Structural Properties}

In order to match the hexagonal lattice of the BN nanotube with the square lattice of the FeO layer it is necessary that
the unit cell of the two lattices commensurate along the tube axis. The lattice parameter along the tube axis of a $\emph{zigzag}$ tube
is  4.35~\AA~(corresponding to 3 times the B-N bond length),  using previous DFT/PBE calculations\cite{BOR2010} on hexagonal BN. The lattice parameter of a FeO square lattice is 4.30~\AA~(which is 2 times the Fe-O bond length). The ratio between the lattice parameters of those two lattices is 0.99. Thus we can assume both structures as matching ones, since errors of 1$\%$ in lattice parameters are imposed in our calculations due to the used approximations described in the Methodology. 

\begin{figure}
 \centering
\includegraphics[scale=0.45,clip=true]{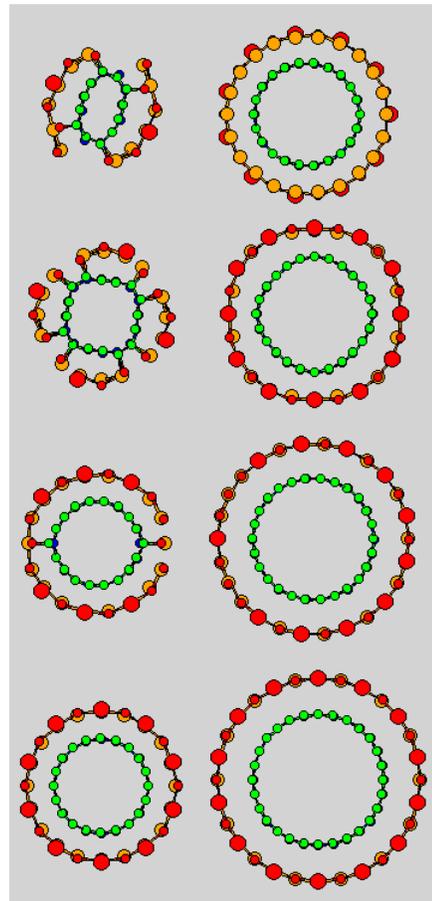}
 \caption{Structure of FeO covered boron nitride (BN) nanotubes. The left column shows, from  top to bottom, 
the structure of nanotubes composed of FeO covering the (6,0), (8,0), (9,0) and (10,0) BN tubes. The right column shows, from top to bottom, nanotubes composed of FeO covering the (11,0), (12,0), (13,0) and (14,0) BN tubes.}
 \label{nanotubos}
\end{figure}

Fig. \ref{nanotubos} shows the cross section of the optimized structure of the studied FeO covered BN nanotubes. 
In the smallest tube cases, i.e., (6,0), (8,0) and (9,0) the large curvature of the tubes impose a large Fe-O bond length in the FeO layer. Since the $\emph{sp}^{2}$ B-N bond is among the hardest bonds in nature, the FeO layer does not
compress the BN tubes. Instead, some Fe-O bonds break up in the FeO layer. Due to the strong interaction between the FeO and the
BN tube,  the FeO  layer covers localized portions of the tube surface  -- which completely changes
the radial shape of the BN tube -- instead of agglomerate in structures with higher coordination number. 
Two diametrically opposed Fe-O  bonds break up in the (6,0) tube  leading to an interesting 
flat tube. Four Fe-O bonds break up in the (8,0) tube  resulting in a square-like cross section. In the case of a (9,0) tube, two neighboring 
Fe-O bonds break up and a slightly flat tube is obtained.      
    
\begin{figure}
 \centering
\includegraphics[scale=0.35,clip=true]{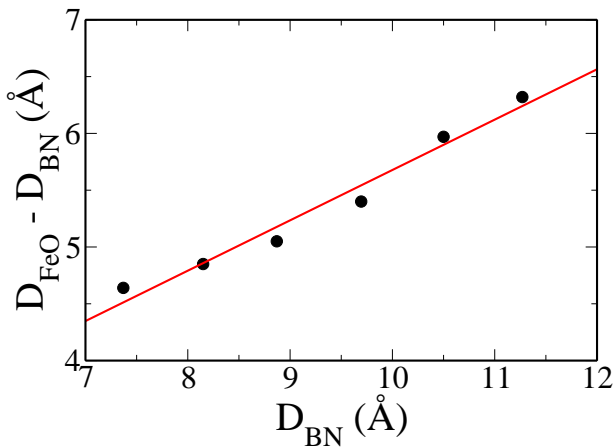}
 \caption{difference between the diameter of the FeO layer ($D_{FeO}$) and the diameter of the BN tube ($D_{BN}$) as a function of $D_{BN}$.
The roughly linear dependence indicates that is harder to compress the Fe-O bond than to increase the distance between the FeO layer and the BN 
nanotubes.}
 \label{diameter}
\end{figure}

In the larger tube cases  (shown in Fig. \ref{nanotubos}), i.e., from (9,0) to (14,0) the increment of the tube diameter leads to a roughly linear increasing in the distance between the FeO layer and the BN tube as can 
be seen in Fig. \ref{diameter}. Such a linear behaviour can be understood if one observes that it is much harder to compress a Fe-O  
bond than to stretch the bond between the BN tube and the FeO layer. In this case it is possible to show that the difference between 
the diameter of the FeO layer ($D_{FeO}$) and the diameter of the BN tube 
($D_{BN}$) is given by:
 
\begin{equation}
D_{FeO}-D_{BN}=D_{BN}\frac{a_{FeO}-a_{BB}}{a_{BB}},
\label{eq1}
\end{equation}
where $a_{FeO}$ is the distance between Fe and O atoms in the FeO layer and $a_{BB}$ is the distance between two B or two N atoms
in the BN tube. Thus, in the cases of large values of $D_{BN}$,  the attractive interaction between the FeO layer and the BN tube must be small,
which accounts against the stability of FeO covered nanotubes. On the other hand, the tube curvature decreases with $D_{BN}$ and so does
the elastic energy associated with it. Therefore, there must be a FeO covered tube
whose formation energy per atom, or equivalently per FeO molecule, is a minimum.

The formation energy ($E_f$) per FeO molecule ($N_{{FeO}}$) can be obtained from total energy calculations as follows:
\begin{equation}
\frac{E_f}{N_{{FeO}}}=\frac{E_{{cover}}-E_{{tube}}}{N_{FeO}}-\mu_{FeO},
\label{eq2}
\end{equation}
where: $E_{{cover}}$ is the first-principles total energy of the FeO covered BN nanotube; $E_{{tube}}$
is the total energy of the BN tube; and $\mu_{{FeO}}$
is the chemical potential of a FeO molecules reservoir. In this work, $\mu_{{FeO}}$ is defined as the total energy
per FeO molecule of a square lattice of FeO. 

\begin{figure}
 \centering
\includegraphics[scale=0.35,clip=true]{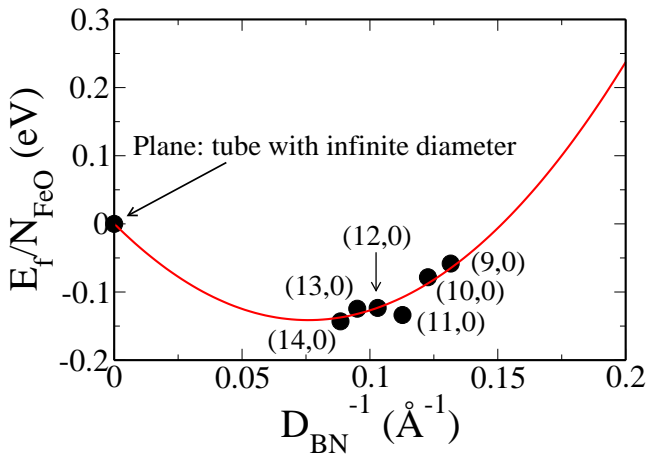}
 \caption{Formation energy divided by the number of FeO molecules, calculated according to Eq. (\ref{eq2}), as a function of the inverse of the BN tube diameter.
The continuous line represents a continuous model, Eq. (\ref{eq3}), which has two contributions: (i) the elastic energy associated 
with the tube curvature and (ii) the attractive interaction between the FeO layer and the BN nanotube.}    
 \label{modelo}
\end{figure}

Figure \ref{modelo} shows the formation energy per FeO molecule of some tubes shown in Fig. \ref{nanotubos} and the formation energy 
of an infinity tube, which is the formation energy per FeO molecule of a plane square lattice of FeO infinitely distant from a sheet of
hexagonal BN. Since the square lattice of FeO is our choice of chemical potential, the infinity tube has null value of formation energy.
The values of formation energies of all studied tubes are negative which shows that it is energetically favorable to
cover a BN nanotube with a FeO layer rather than to form an isolated FeO layer. 
The coordination number of each atom in the FeO layer, 
which is four, is larger than the average coordination number of very small FeO nanoparticles (for instance, the average coordination number of
a body centered cubic FeO nanoparticle with 32 atoms is 3.75). This makes the FeO surface energetically more stable than such small nanoparticles 
(the formation energy of the FeO nanoparticle with 32 atoms is 0.75~eV). Then, it is energetically favorable for the FeO molecules
to cover the BN tube rather than to aggregate forming small nanoparticles,  suggesting that the FeO bulk would not be formed until the BN tubes were covered.      

Figure \ref{modelo} shows that the formation energy tends to decrease as much as the tube grows from (9,0) to (14,0). Since
the formation energy of a tube with infinite diameter (flat surface) is zero, the values of energy must start to increase with $D_{BN}$ after some point of 
minimal energy. In order to estimate the minimal energy point we propose a continuous model which takes into account the elastic
energy per FeO molecule, or BN pair, associated with the tube curvature and the interaction between the FeO layer and the BN tube.

The energy associated with the tube curvature is due the distortions imposed to the B-N and Fe-O bonds to form the tube,
which scales with $C/D_{BN}^{2}$, where $C$ is a constant. \cite{B2010,BMC2006} The interaction between the FeO layer and the BN tube $E_{(FeO-BN)}$
 depends on the difference between 
$D_{FeO}$ and $D_{BN}$, which is proportional to $D_{BN}$ in case of tubes larger than (9,0) [see Eq. (\ref{eq1})].
Such an interaction must vanish as $D_{BN}$ $\rightarrow \infty$  and blows up as $D_{BN}$  $\rightarrow 0$.\footnote{In fact, the PBE functional can describe only 
the short-range part of van der Waals force, it is unable to describe the $-C_{6}$/$R^{2}$ behaviour of long-range interactions,
instead it predicts a exponential decay at long-range 
(see Ref. \cite{RPC2005})}Thus, 
the interaction between the FeO layer and the BN tube can be written as follows: 
  
\begin{equation} 
\frac{E_{form}}{N_{FeO}}= C_{0}+\frac{C_{1}}{D_{BN}}+\frac{C_{2}}{D_{BN}^{2}},
\label{eq3}
\end{equation}
where $C_{0}$ is the formation energy per FeO molecule of a tube with infinite diameter. $C_{2}$ includes the contributions of
both the elastic energy associated with the tube curvature and the interaction between the tube and the FeO layer. Here we have neglected terms with order higher than 3 since they are small in comparison to the energy associated with the curvature
$D_{BN}^{2}$. Therefore, Eq. (\ref{eq3}) must be a reasonable estimation for the formation energy per FeO molecule of FeO covered BN nanotubes.
      
The continuous line in Fig. \ref{modelo} represents the proposed continuous model. According to such a model the minimal value of
formation energy occurs at $D_{BN}$=13.22 \AA.      

\section{Electronic Structure}

 The band structure of large FeO covered nanotubes (large enough to allow us to neglect the curvature effects) 
must be a superposition of the band structure of the hexagonal BN layer and the band structure of
a square FeO layer since the distance between the FeO layer and the BN nanotubes increases with $D_{BN}$. 
BN structures (including hexagonal lattices and nanotubes) present a large energy band gap of
roughly 5~eV. Therefore, electronic states near the fermi level should be due to the FeO layer in such
large tubes. Thus, as a first step to understand the band structure of the FeO covered BN nanotubes, we have
investigated the bands structure of the FeO square lattice. 

\begin{figure*}
 \centering
\includegraphics[scale=0.4,clip=true]{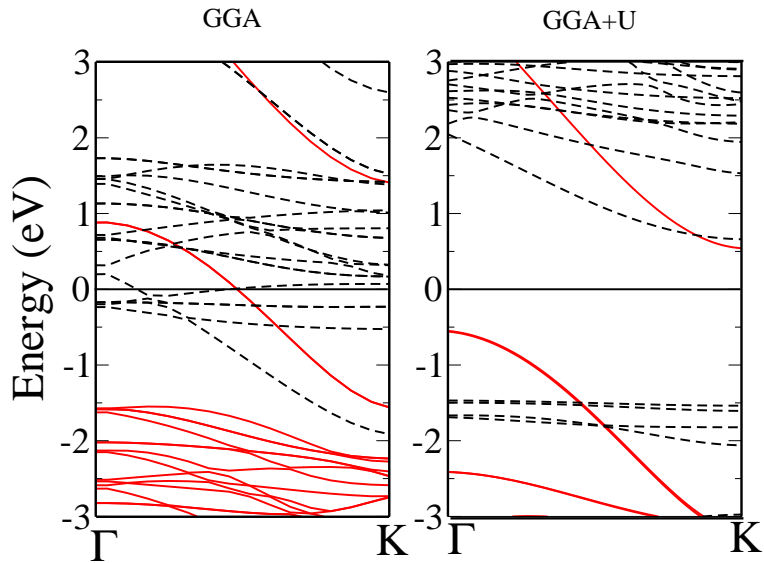}
 \caption{Bands structure of a FeO square lattice. Left panel: PBE calculations. Right panel
PBE+U calculations. Continuous red lines: majority spin states. Dashed black lines: minority spin states.
The inclusion of the Hubbard term leads to a integer occupation of Fe $3d$ orbitals and
more localized Fe orbitals. As a result the low-dispersion bands nearby the fermi level disappear which
leads to a 1.2 band gap. The fermi level was set to zero.}
 \label{superbands}
\end{figure*}

Similarly to the case of FeO bulk, the GGA approach predicts a metallic behavior for the FeO layer while the GGA+U predicts
a 1.2 eV band gap for FeO layer (see Fig. \ref{superbands}). The GGA bands structure shows several low dispersion bands (dashed lines)
nearby the fermi level which are associated with the over-delocalized Fe $3d$ orbitals. The
inclusion of the Hubbard term leads to an integer occupation of Fe $3d$ orbitals and  
more localized Fe orbitals. As a result the low-dispersion bands nearby the fermi level disappear resulting in
 a 1.2 eV band gap. 

\begin{figure*}
 \centering
\includegraphics[scale=0.45,clip=true]{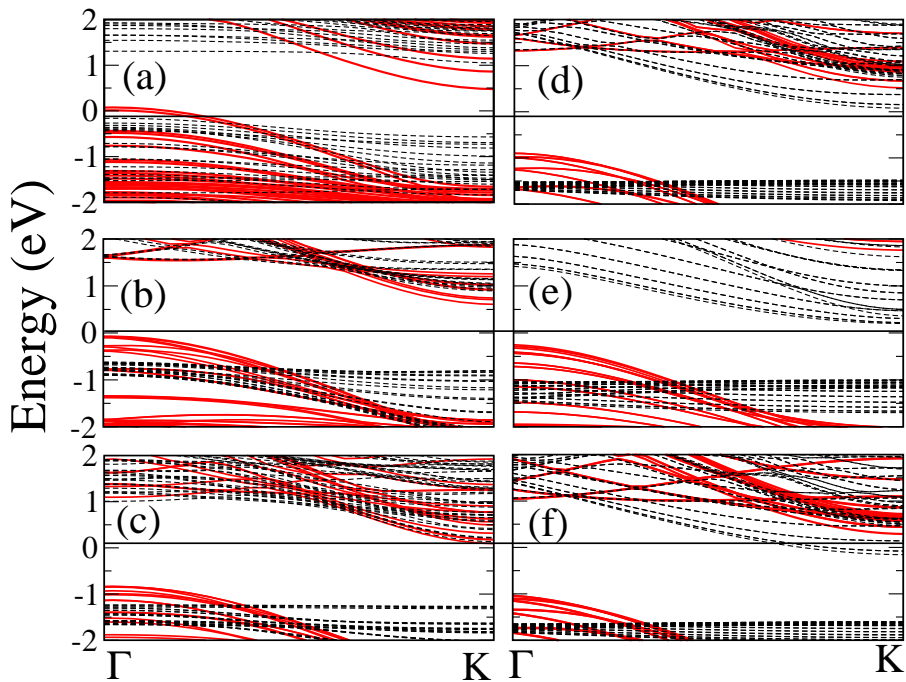}
 \caption{Bands structure of FeO covered BN nanotubes shown in Fig. \ref{nanotubos}: (a)
 (10,0) FeO covered BN nanotube; (b) (11,0) FeO covered BN nanotube; (c)(12,0) FeO covered BN nanotube;
(d) (14,0) FeO covered BN nanotube. (e) Bands structure of a (14,0) FeO layer. (f) Bands structure
of a (14,0) FeO covered BN nanotube charged with -$e$/4.3\AA. 
Continuous red lines: majority spin states. Dashed black lines: minority spin states. The fermi
level was set to zero.
  }
 \label{bandatubo}
\end{figure*}

Although some qualitative information about the 1D energy bands of 
large studied FeO covered nanotubes can be obtained by \emph{slicing} the energy dispersion of the FeO layer
(in the directions expressed by $D_{BN} k_{y}=2\pi n$, where $n$ is a integer), it is not possible
to obtain from the energy dispersion of the FeO layer quantitative information about the studied tubes.
For instance, by \emph{slicing} the energy dispersion of the FeO layer, we found that the bottom of the conduction
band of a (10,0) should be below the lowest unoccupied majority spin states while the bottom of the conduction 
band of a (14,0) should be bellow minority spin states which is in agreement with the calculated energy dispersion
for such tubes  (see Fig. \ref{bandatubo}). On the other hand, by \emph{slicing} the energy dispersion of the FeO layer we expect the band gap
increases in as much as the diameter of the tube decreases, however, 
the band gap decreases with the diameter in as much as the tube size decreases from (14,0)
up to (10,0) as shown in Fig. \ref{bandatubo}. The (10,0) FeO covered BN nanotube presents a band gap (0.5~eV) much smaller than the
value expected by \emph{slicing} the energy dispersion of the FeO layer (1.7~eV). Such
differences are due the curvature and the interaction of FeO layer with the BN tube.
The effect of the curvature can be investigated by performing calculations on pure FeO tube with the same geometry of
FeO covered BN nanotubes but without the inner BN nanotube. 
For a pure (14,0) FeO layer we have found a band gap of 0.5~eV which is much smaller than the value 
obtained by \emph{slicing} the energy dispersion of the FeO layer (1.4~eV). Due to the interaction
with the BN nanotube the band gap of FeO layer increases to 0.9~eV which shows that both
curvature and interaction with the BN tube are determinant for the electronic properties. 

The curvature and the interaction of the FeO with the BN tubes lead to interesting transport properties. The
(14,0) FeO covered BN nanotube  is a semi-half-metal which becomes half-metal upon electron doping 
(see Fig. \ref{bandatubo}). As it is possible to see in Fig. \ref{bandatubo},
the bottom of the conduction band of the neutral (14,0) tube is 0.5~eV bellow the first unoccupied minority spin states, 
therefore, upon electron doping it is possible to change the Fermi level to 
allow the conduction of electrons only through majority spin states. Since the top of valence band
of the (11,0) tube is 0.6~eV above the lowest occupied majority spin states, the (11,0) FeO
covered BN nanotube should become a half-metal upon hole doping. The (10,0) nanotube is a intrinsic half-metal,
the Fermi level cross the highest bands with negative curvature as it is possible to see in Fig. \ref{bandatubo}.

\section{Conclusions}

In summary,  we have applied fist-principles formalism to investigate the effect of the tube diameter on the stability and on the electron transport properties of FeO covered BN nanotubes.  

Regarding  structural properties, we observed that the increment
of the tube diameter leads to a roughly linear increasing in the distance between the FeO layer and the BN tube (see Fig.  \ref{diameter}). Next, we propose a continuous model for the FeO covered BN nanotubes formation energy which predicts that FeO covered BN tubes with diameter of roughly 13 \AA~ are the most stable ones.

For the electronic structure we have found a rich variety of electron transport properties for this system: (i) the (6,0) tube is semiconductor; (ii) the (10,0) is a half-metal; (iii) the (11,0) is a semi-half-metal which could become a half-metal if doped with holes; (iv) the (14,0) is a semi-half-metal which become a half-metal if doped with electrons. 
Similarly to the case of FeO bulk, the GGA functional predicts predicts a metallic behavior for a FeO layer while the GGA+U predicts a semiconductor behavior.   
  
\section*{Acknowledgements}

This work was partially supported by the Brazilian science
agencies CNPq and FAPEMIG.


\bibliographystyle{aip}
 \bibliography{abinitio}

\end{document}